\documentclass[twocolumn,showpacs,preprintnumbers,amsmath,amssymb]{revtex4}
\usepackage{graphics}% Include figure files
\usepackage{epsfig}
\usepackage{bm}% bold math
\begin{document}
%\preprint{Krenke01}
\title{Effects of hydrostatic pressure on the magnetism and martensitic
transition of Ni-Mn-In magnetic superelastic alloys}

\author{Llu\'{i}s Ma\~nosa, Xavier Moya,  Antoni Planes}
\affiliation{Facultat de F\'isica, Departament d'Estructura i
Constituents de la Mat\`eria, Universitat de Barcelona, Diagonal
647, E-08028 Barcelona, Catalonia, Spain}

\author{Oliver Gutfleisch, Julia Lyubina}

\affiliation{IFW-Dresden. Institute of Metallic Materials. P.O.
Box 270116. D-01171 Dresden, Germany}

\author{Maria Barrio, Josep-Llu\'{\i}s Tamarit}

\affiliation{Departament de F\'{\i}sica i Enginyeria Nuclear,
ETSEIB, Universitat Polit\`ecnica de Catalunya, Diagonal 647,
08028-Barcelona, Catalonia, Spain}

\author{Seda Aksoy, Thorsten Krenke\cite{AdressKrenke}, Mehmet Acet}

\affiliation{Experimentalphysik, Universit\"at Duisburg-Essen,
D-47048 Duisburg, Germany}

\date{\today}

\begin{abstract}
We report magnetization and differential thermal analysis
measurements as a function of pressure accross the martensitic
transition in magnetically superelastic Ni-Mn-In alloys. It is
found that the properties of the martensitic transformation are
significantly affected by the application of pressure. All
transition temperatures shift to higher values with increasing
pressure. The largest rate of temperature shift with pressure has
been found for Ni$_{50}$Mn$_{34}$In$_{16}$ as a consequence of its
small entropy change at the transition. Such a strong pressure
dependence of the transition temperature opens up the possibility
of inducing the martensitic transition by applying relatively low
hydrostatic pressures.

\end{abstract}


%\pacs{81.30.Kf, 75.50.En, 75.50.Cc }

\maketitle

The Ni-Mn based Heusler compounds with compositions close to the
stoichiometric Ni$_2$Mn$X$ (with $X$ being  group IIIA-VA
elements) have been shown to exhibit many functional properties
such as magnetic shape memory \cite{Ullakko96}, magnetic
superelasticity \cite{Krenke07a}, magnetocaloric effects
\cite{Krenke05a} and magnetoresistance \cite{Sharma06}, which
derive from the coupling between the martensitic transition and
the magnetic order. In this family of alloys, magnetic moments are
mainly confined to the Mn atoms. The exchange interaction between
magnetic moments is long range and oscillatory, and is mediated by
the conduction electrons. As a consequence, the magnetic
properties of these alloys are sensitive to the distance between
neighboring Mn atoms, and, indeed, different magnetic behavior has
been reported for alloys with different $X$ element. At the
martensitic transition, the change in the lattice cell modifies
the distance between Mn-atoms which can lead to antiferromagnetic
interactions. Antiferromagnetic interactions are expected to be
present in  off-stoichiometric Ni-Mn-$X$ compounds with $X$ as Ga
\cite{Enkovaara03}, Sn \cite{Krenke05b}, In \cite{Krenke06} and Sb
\cite{Khan07}.

In the present paper, we investigate the effect of hydrostatic
pressure on Ni-Mn-In magnetic shape memory alloys. The application
of pressure modifies the distance between Mn atoms thereby affecting the magnetic exchange. The
relative stability between the high temperature cubic phase
and the low temperature martensitic phase is also be affected by
pressure.

Two samples were prepared by arc melting pure metals under argon
atmosphere. They were then annealed at 1073 K for 2 hours and
quenched in ice-water. The compositions of the alloys were
determined by energy dispersive x-ray analysis to be
Ni$_{50.0}$Mn$_{34.0}$In$_{16.0}$ and
Ni$_{49.5}$Mn$_{35.5}$In$_{15.0}$. Magnetization measurements were
performed using a superconducting quantum interference device
magnetometer equipped with pressure cell in fields up to 5 T in
the temperature range 4 - 340 K and for pressures up to 10 kbar.
High-pressure differential thermal analysis (HP-DTA) was carried
out in a calorimeter capable of operating in the temperature and
pressure ranges 183 - 473 K and 0 - 3 kbar respectively. The
calorimeter is similar to that described in \cite{Wurflinger75}.
Powder samples were mixed with an inert perfluorinated liquid
(Galden®, from Bioblock Scientifics) before they were hermetically
sealed in order to ensure pressure transmission. Thermal curves
were recorded as a function of temperature for selected pressure
values. HP-DTA scans were run on heating and cooling at 1 K
min$^{-1}$ rates.

\begin{figure}
\includegraphics[width=6.9cm]{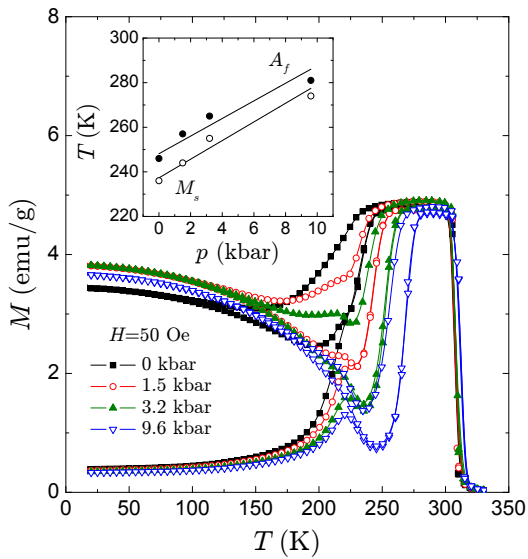}
\caption{\label{MvsT} (Color online) (a) Low field magnetization
versus temperature curves for selected applied pressures. The
inset shows the martensitic start transition temperature ($M_s$)
and austenite finish temperature ($A_f$) as a function of the
applied pressure.}
\end{figure}

Figure \ref{MvsT} shows the temperature dependence of the
magnetization in a low external magnetic field of $H=50$ Oe at
selected applied pressures for Ni$_{50.0}$Mn$_{34.0}$In$_{16.0}$.
Data have been taken in a zero-field-cooled (ZFC), field-cooled
(FC), and field-heated (FH) sequence. Results for $p=0$ agree with
those previously reported \cite{Krenke06}. On cooling, the cubic
phase orders ferromagnetically at $T_{C}^{A}\simeq 310$ K which
causes a sharp increase in the magnetization. At a lower
temperature $M_s$, the sample transforms to the martensitic phase,
and there is a sharp drop in the magnetization. Upon further
cooling the magnetization rises again, reflecting the increase in
ferromagnetic order of the martensite at the Curie point of the
martensitic phase $T_{C}^{M}$. The hysteresis in the FC and FH
curves is a consequence of the first order character of the
martensitic transition, while the splitting between ZFC and FH
curves is associated with the presence of low temperature
anisotropy together with any possible antiferromagnetic components
existing in the martensitic state.

\begin{figure}
\includegraphics[width=6.9cm]{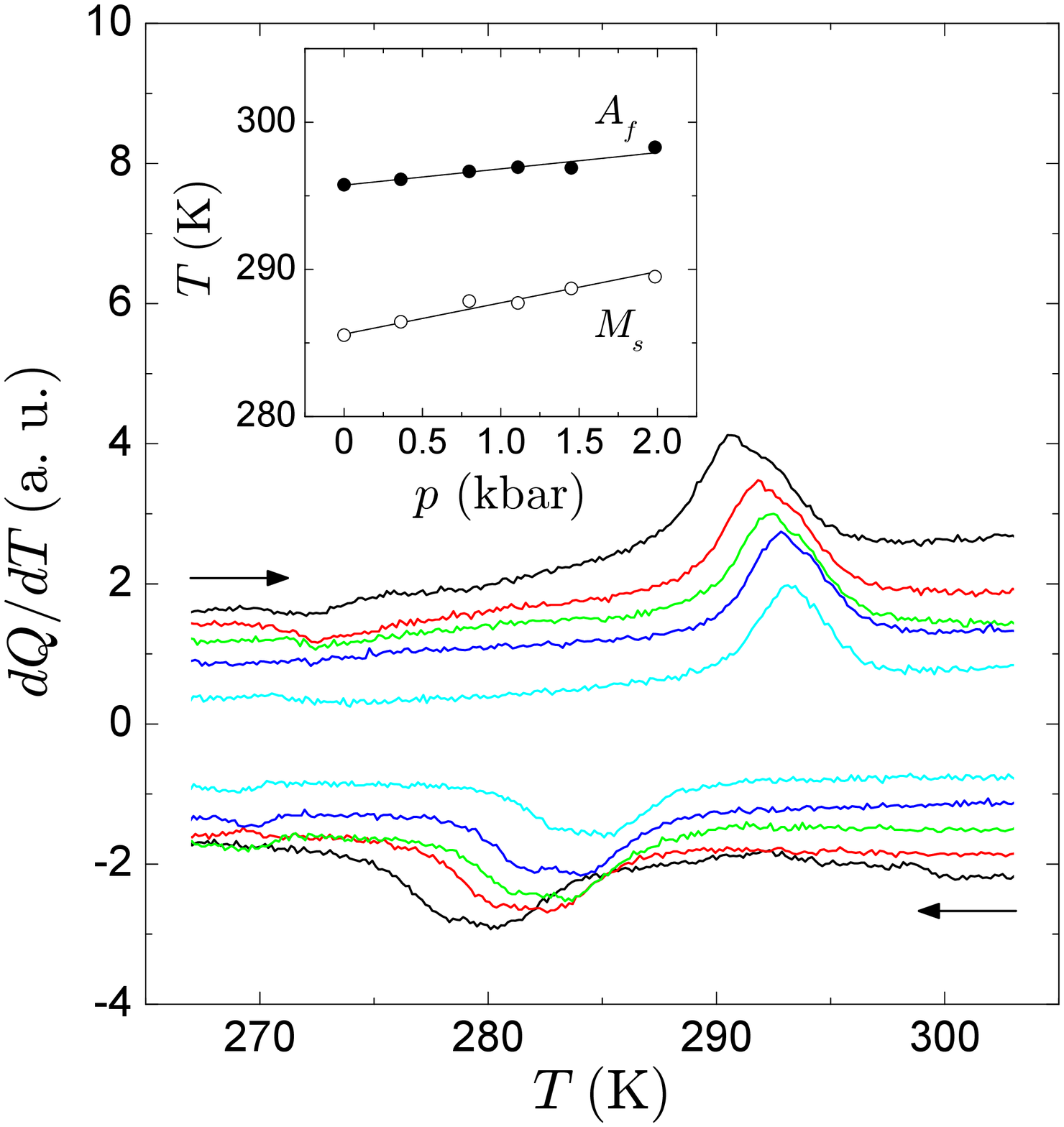}
\caption{\label{dQdTvsT} (Color online) (a) DTA curves for
selected values of applied pressure. From top to bottom (heating)
and bottom to top (cooling) the applied pressures are: 0 kbar,
0.36 kbar, 0.80 kbar, 1.11 kbar and 1.45 kbar. The inset shows the
martensitic start transition temperature ($M_s$) and austenite
finish temperature ($A_f$) as a function of the applied pressure.}
\end{figure}

The application of pressure has little effect on the magnetic
behavior of the high temperature cubic phase. Magnetization values
below $T_{C}^{A}$ are coincident within experimental error for all
applied pressures. The $T_{C}^{A}$ marginally increases with
pressure, in agreement with data reported for related Heusler
alloys \cite{Kaneko81,Kyuji97,Albertini07} and are consistent with
the predictions of first principles calculations
\cite{Sasioglu05}. Also, below about 150 K, well in the
martensitic state, the temperature behavior of the magnetization
remains nearly the same at all pressures. However, pressure has a
significant effect on the magnetic behaviour in the temperature
region where the austenitic and martensitic phases coexist. All
characteristic temperatures associated with the martensitic
transition shift to higher values as the pressure is increased.
The inset in Fig. \ref{MvsT} shows the pressure dependence of
$M_s$ and $A_f$ which exhibit similar behaviour with pressure.
Another feature is that  the change in the magnetization between
martensite and austenite becomes larger with increasing pressure.
Such an increase is consistent with the fact that the martensitic
transition is shifted to higher temperatures, and along with this,
the ferromagnetic order of the martensite decreases as the
temperature increases. Also, application of pressure is expected
to enhance any antiferromagnetic exchange present in the
martensitic phase.

To gain further information on pressure effects on the martensitic
transition, we have performed DTA measurements under pressure. The
thermal curves for powder Ni$_{49.5}$Mn$_{35.5}$In$_{15.0}$ at
selected hydrostatic pressures are shown in Fig. \ref{dQdTvsT}.
The endothermal and exothermal peaks corresponding to the reverse
and forward transitions on heating and cooling respectively are
visible on the curves. Application of pressure does not
significantly alter the shape of the thermal peak. For this
sample, both forward and reverse transitions also shift towards
higher temperatures as the pressure increases. The rate of shift
in the transition temperatures in
Ni$_{49.5}$Mn$_{35.5}$In$_{15.0}$ $dT/dp \approx 2$ K kbar$^{-1}$
is lower than in Ni$_{50}$Mn$_{34}$In$_{16.0}$ with $dT/dp \approx
4$ K kbar$^{-1}$.

For first order phase transitions, the shift in the transition
temperatures is accounted for by the Clausius-Clapeyron equation
$dT/dp = \Delta v/\Delta S$, where $\Delta S$ and $\Delta v$ are
respectively the entropy and volume changes at the phase
transition. Complementary differential scanning calorimetry
measurements have been performed which give $\Delta S=2.17 $ J
mol$^{-1}$ K$^{-1}$ for Ni$_{49.5}$Mn$_{35.5}$In$_{15}$ and
$\Delta S=0.53 $ J mol$^{-1}$ K$^{-1}$ for
Ni$_{50}$Mn$_{34}$In$_{16}$. By using a molar volume of  $v=8.9$
m$^3$ mol$^{-1}$ for 15.0 at\% In and $v=9.2$ m$^3$ mol$^{-1}$ for
16 at\% In computed from X-ray data, we obtain the relative volume
changes at the martensitic transition of 0.6 \% and 0.3 \% for
Ni$_{49.5}$Mn$_{35.5}$In$_{15.0}$ and
Ni$_{50.0}$Mn$_{34.0}$In$_{16.0}$ respectively. The larger shift
in the transition temperatures found for the 16 at\% sample is due
to its lower value of $\Delta S$ at the transition.

\begin{figure}
\includegraphics[width=6.9cm]{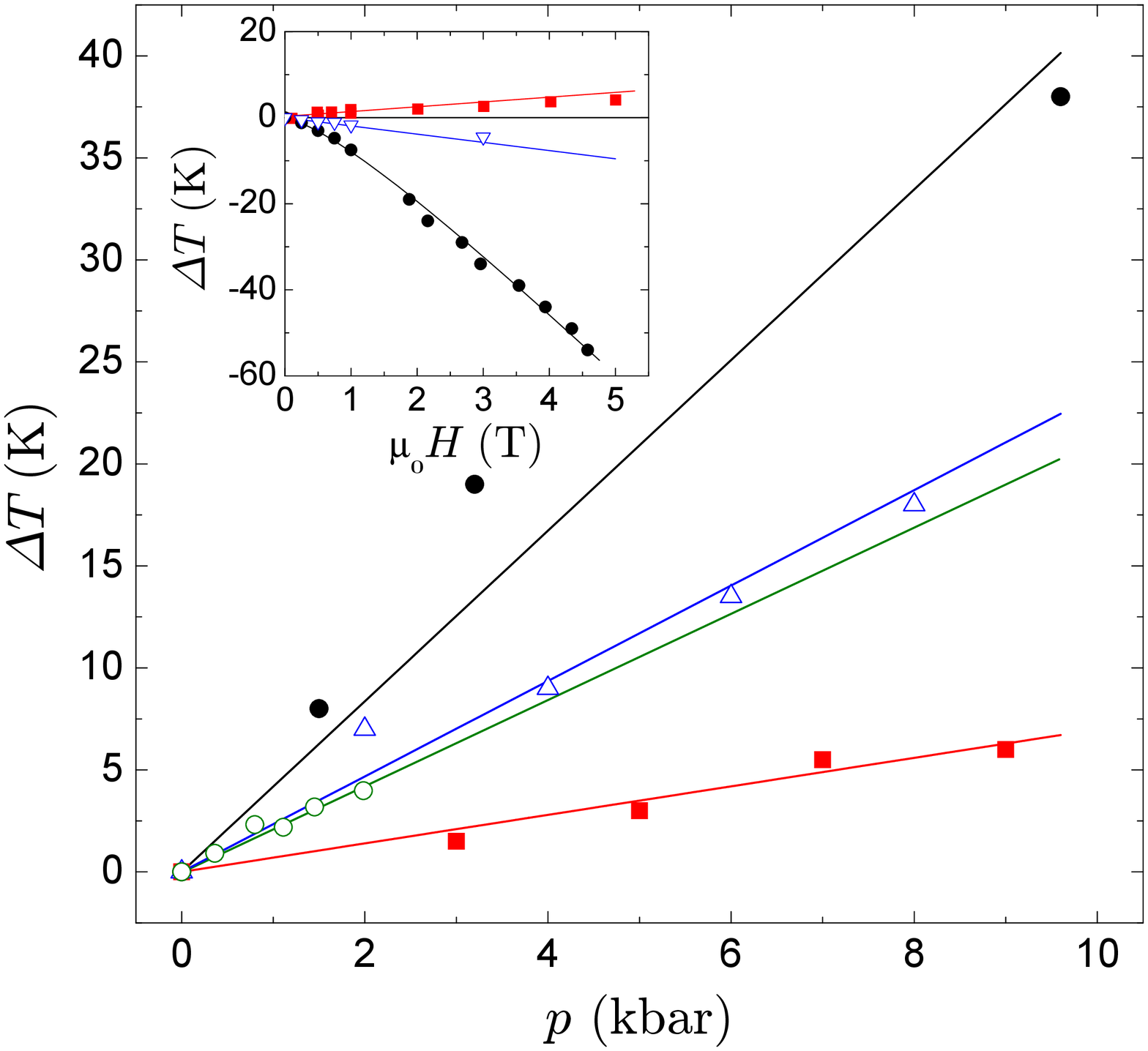}
\caption{\label{Tvsp&H} (Color online) Shift with pressure of the
martensitic transition temperature for
Ni$_{53.5}$Mn$_{23}$Ga$_{23.5}$ (squares) \cite{Kim06},
Ni$_{50}$Mn$_{36}$Sn$_{14}$ (up triangles) \cite{Yasuda07},
Ni$_{49.5}$Mn$_{35.5}$In$_{15}$ (open circles) and
Ni$_{50}$Mn$_{34}$In$_{16}$ (solid circles). Lines are fits to the
data. The inset shows the shift in the transition temperature with
magnetic field. Data correspond to the same samples than for the
pressure dependence but for Ni-Mn-Sn, the line corresponds to
Ni$_{50}$Mn$_{36}$Sn$_{14}$ \cite{Yasuda07} while down triangles
stand for Ni$_{50}$Mn$_{35}$Sn$_{15}$ \cite{Moya07}.}
\end{figure}

Figure \ref{Tvsp&H} compares the shift in the martensitic
transition temperature found for Ni-Mn-In alloys to those reported
for several Ni-Mn-$X$ shape memory alloys \cite{footnote}. In all
cases, the martensitic transition temperature increases with
increasing pressure as a consequence of the lower volume of the
martensitic phase with respect to the cubic phase. For Ni-Mn-Ga
alloys, the rate of change is smaller than for Ni-Mn-Sn and
Ni-Mn-In. Such a difference is due to the fact that for this
alloy, the relative volume change  is smaller \cite{Kim06} than
 for Ni-Mn-Sn  \cite{Yasuda07} and Ni-Mn-In \cite{Aksoy07}.
In magnetic shape memory alloys, the coupling between magnetism
and structure results in a magnetic field dependence of the
structural transition. The inset in Fig. \ref{Tvsp&H} shows the
magnetic field dependence of the martensitic transition
temperature for the same samples for which data are shown in Fig.
\ref{Tvsp&H}. For Ni-Mn-Ga, there is a slight increase in the
transition temperature with increasing magnetic field, while in
Ni-Mn-Sn and Ni-Mn-In, the transition temperatures decrease with
increasing field. Such different behavior is due to the fact that
the saturation magnetization of the martensitic phase is larger
than that of the austenite in Ni-Mn-Ga while for Ni-Mn-Sn and
Ni-Mn-In, the martensitic state has a lower magnetization than the
austenitic state. We note that the rate of change in the
transition temperature with both pressure and magnetic field for
the sample with 16 at$\%$ In is much larger than in other
Ni-Mn-$X$ alloys. Such behavior is due to the lower entropy change
in this alloy as compared to the entropy change in the other
alloys. Therefore, for this alloy, it is easier to induce the
martensitic transition by applying moderate hydrostatic pressure
or magnetic field as opposed to the other compounds. This feature
opens up a broad range of possible applications of the functional
properties of this alloy, such as magnetic superelasticity,
caloric effects, magnetoresistance, etc., associated with a
pressure or magnetic field induced martensitic transition.

This work was supported by CICyT (Spain), projects MAT2007-62100
and FIS2005-00975, and by Deutsche Forschungsgemeinschaft projects
RI 932/4-1 and SPP 1239. XM acknowledges support from DGICyT
(Spain).


\begin{thebibliography}{99}
\bibitem{footnote} Notice that because all characteristic temperatures shift
similarly with pressure, the relative change with pressure of any
of them can be used to characterize the shift of the transition
with pressure.
\end{thebibliography}
\end{document}